\documentclass[aps,prl,twocolumn,amsfonts,showpacs,floatfix]{revtex4}
\usepackage{epsfig}
\usepackage{times}

\def\QED{\leavevmode\unskip\penalty9999 \hbox{}\nobreak\hfill
     \quad\hbox{\leavevmode  \hbox to.77778em{%
               \hfil\vrule   \vbox to.675em%
               {\hrule width.6em\vfil\hrule}\vrule\hfil}}
     \par\vskip3pt}
\def\qed{\leavevmode\unskip\penalty9999 \hbox{}\nobreak\hfill
     \quad\hbox{\leavevmode  \hbox to.77778em{%
               \hfil\vrule   \vbox to.675em%
               {\hrule width.6em\vfil\hrule}\vrule\hfil}}
     \par\vskip3pt}
\newcommand{\trace}{\mathop{\rm Tr}\nolimits}
\newcommand{\text}{\mbox}

\newcommand{\identity}{\openone}
\newcommand{\be}{\begin{equation}}
\newcommand{\ee}{\end{equation}}
\newcommand{\bea}{\begin{eqnarray}}
\newcommand{\eea}{\end{eqnarray}}
\newcommand{\beas}{\begin{eqnarray*}}
\newcommand{\eeas}{\end{eqnarray*}}

\begin{document}

\title{The entanglement cost under operations
preserving the positivity of partial transpose}

\author{K.\ Audenaert}
\author{M.B.\ Plenio}
\author{J.\ Eisert}
\affiliation{QOLS, Blackett Laboratory, Imperial College
of Science, Technology and Medicine, London,
SW7 2BW, UK}


\date{\today}
\begin{abstract}
We study the entanglement cost under quantum operations preserving
the positivity of the partial transpose (PPT-operations). We
demonstrate that this cost is directly related to the logarithmic
negativity, thereby providing the operational interpretation for
this easily computable entanglement measure. As examples we
discuss general Werner states and arbitrary bi-partite Gaussian
states. Equipped with this result we then prove that for the
anti-symmetric Werner state PPT-cost and PPT-entanglement of
distillation coincide giving the first example of a truly mixed
state for which entanglement manipulation is asymptotically
reversible.
\end{abstract}

\pacs{03.67.Hk}

\maketitle

The theory of quantum entanglement is closely intertwined with the
study of those quantum operations that can be locally implemented
in quantum systems consisting of more than one subsystem. If one
also allows for the classical transmission of outcomes of local
measurements, then one arrives at the set of local quantum
operations with classical communication (LQCC). This set of
quantum operations reflects on the one hand the typical physical
restrictions imposed by the setup of many  basic applications of
quantum information theory \cite{Introduction}. On the other hand,
the very notion of entanglement is intimately related to this set
of operations. For example, one calls a quantum state entangled if
it can not be prepared using LQCC, in contrast to so-called
separable states.

The study of entanglement manipulation is concerned with the
transformation from one entangled state to another by means of
LQCC. Not very surprisingly, one finds that for any finite number
of identically prepared quantum systems such manipulation of
entanglement under LQCC is generally irreversible, both for pure
and mixed states. In fact, the pure state-case can be most easily
assessed, as powerful necessary and sufficient criteria for the
interconvertibility of entangled states have been found
\cite{pureconditions}.
In the asymptotic limit of infinitely many identical copies of a
pure state, in contrast, pure bi-partite entanglement can be
interconverted reversibly \cite{entropy}. This statement can also
be cast in the language of entanglement measures. These are
functions of a quantum state that cannot increase under a given
set of operations (e.g.\ LQCC). Entanglement measures are useful
mathematical and conceptual tools and several such measures have
been suggested, most notably the entanglement of formation
\cite{entform}, the distillable entanglement \cite{entform,Rains
99} and the relative entropy of entanglement \cite{Rains 99,Vedral
PRK 97,Audenaert EJPVD01,Audenaert DVW02}. The distillable
entanglement is, essentially, defined as the asymptotic number of
pure maximally entangled states that can be extracted via LQCC
from a set of identically prepared quantum systems. Analogously,
the entanglement cost is defined as the asymptotic number of pure
maximally entangled state that are required to create a given,
possibly mixed state. The asymptotic reversibility of pure state
entanglement is then equivalent to the statement that the
entanglement cost and the entanglement of distillation are in fact
equal for pure states. Then a single number,  the von Neumann
entropy of a subsystem, uniquely quantifies the degree of
entanglement. For mixed states, however, this asymptotic
reversibility under LQCC operations is lost again. Examples have
been found for which the entanglement cost and the entanglement of
distillation are provably different \cite{Cirac 02,Horodecki
SS02}.

The study of general asymptotic entanglement manipulation -- while
formally being at the roots of a theory of entanglement -- is
complicated by the fact that the characterization of LQCC
themselves is far from being well-understood. However, there is a
closely related set of operations that can be much more easily
characterized, namely that of PPT-preserving operations
(PPT-operations in brief). These operations are defined as those
that map any state which has positive partial transpose into
another state with positive partial transpose. PPT-operations are
more powerful than LQCC operations as they allow, for example, the
creation of any bound entangled state from a product state and
ensure the distillability of any NPT-state, i.e.\ any state that
cannot be created by PPT-operations \cite{Eggeling VWW01}. As a
consequence, the set of states decomposes into two subsets, the
PPT-states (non-distillable) and the states that are distillable
under PPT-operations. This provides a significant simplification
of the entanglement structure under PPT-operations as compared to
that under LQCC operations where at least three classes of states,
disentangled, bound entangled and distillable are known. Indeed,
the results presented in the following point towards the
possibility that the structure of entanglement under
PPT-operations is even simpler, namely that PPT-entanglement cost
and PPT-distillable entanglement may be equal or, in other words,
that entanglement may be asymptotically reversible under
PPT-operations.

We start by summarizing the main results of this paper. Firstly,
we prove that the PPT-entanglement cost for the {\em exact}
preparation of a large class of quantum states under
PPT-operations is given by the logarithmic negativity \cite{Vidal
W02}, thus providing an operational meaning to the logarithmic
negativity. Secondly, we employ this result to show that the
PPT-entanglement cost of the anti-symmetric Werner state in any
dimension is given by the logarithmic negativity thereby
demonstrating that the PPT-cost is equal to the PPT-entanglement
of distillation for this state. This is the first example of a
truly mixed state for which the entanglement manipulations have
been proven to be asymptotically reversible. We end this work with
a discussion of the implications that this result has, including
the possibility of the reversibility of PPT-entanglement
manipulations for all states.

Before we state and formally prove our results we introduce a few
basic concepts, including the definitions of the PPT-entanglement
of distillation and the PPT-entanglement cost. We introduce the
notation (following Rains \cite{Rains 00}) where $\Psi$ denotes a
trace preserving completely positive map and $\Phi(K)$ is the
density operator corresponding to the maximally entangled state
vector in $K$ dimensions, i.e.\
$\Phi(K)=|\psi^+\rangle\langle\psi^+|$ with $|\psi^+\rangle =
\sum_{i=1}^K |ii\rangle/\sqrt{K}$. The PPT-distillable
entanglement is defined as
\begin{displaymath}
    D_{ppt}(\rho) = \sup\{ r: \lim_{n\rightarrow\infty}
    \sup_{\Psi} {\rm tr}[\Psi(\rho^{\otimes n})\Phi(2^{rn})] = 1
    \}.
\end{displaymath}
For the PPT-entanglement cost of a quantum state $\rho$ we study
two definitions that correspond to different requirements in the
preparation of the state $\rho$. The standard definition of the
PPT-entanglement cost $C_{ppt}$ requires that the quality of the
approximation of the state $\rho^{\otimes n}$ by $\Psi(\Phi(K))$
becomes progressively better and converges in the asymptotic limit
under the trace norm, or, formally
\begin{displaymath}
    C_{ppt}(\rho) = \inf\{ r: \lim_{n\rightarrow\infty}
    \inf_{\Psi} {\rm tr}|\rho^{\otimes n}-\Psi(\Phi(2^{rn}))| = 0 \} .
\end{displaymath}

However, for a more restrictive definition one requires the {\em
exact} preparation of any finite number of copies of the state and
not just the asymptotically exact preparation. This quantity,
$E_{ppt}$, which will generally be larger than $C_{ppt}$, reads
formally as
\begin{displaymath}
    E_{ppt}(\rho) = \lim_{n\rightarrow\infty} \inf\{ r_n: \inf_{\Psi}
    {\rm tr}|\rho^{\otimes n}-\Psi(\Phi(2^{r_n n}))| = 0 \}.
\end{displaymath}
This quantity will later be related to the logarithmic negativity,
which was defined in \cite{Vidal W02} as
\begin{displaymath}
    LN(\rho)=\log_2 {\rm tr} |\rho^{\Gamma}|,
\end{displaymath}
where $\rho^{\Gamma}$ stands for the partial transpose of the
density operator $\rho$. While the negativity ${\rm tr}
|\rho^{\Gamma}|$ is an entanglement monotone (including convexity)
\cite{Vidal W02,Thesis}, the logarithmic negativity is a monotone
only under non-selective PPT-preserving operations. Apart from the
partial transposition of a density operator another important
quantity for the following will be the so-called bi-negativity
$|\rho^{\Gamma}|^{\Gamma}$ \cite{Audenaert DVW02}. While its
physical interpretation is not yet properly understood, it plays a
significant role in the following theorems and has proven to be a
useful concept in investigations of entanglement manipulations
\cite{Audenaert DVW02}.
After these basic definitions we are now in a position to present
and prove the first theorem concerning the PPT-entanglement cost.

\smallskip
\noindent {\bf Theorem:} {\it The PPT-entanglement cost
$E_{ppt}(\rho)$ for the exact preparation of the state $\rho$
satisfies $$
    \log_2 {\rm tr} |\rho^{\Gamma}| \le E_{ppt}(\rho) \le \log_2 Z(\rho)
$$ where $$
    Z(\rho) = {\rm tr} |\rho^{\Gamma}| + \dim(\rho) \max
    (0,-\lambda_{min}(|\rho^{\Gamma}|^{\Gamma}).
$$ }

{\it Proof:} The lower bound follows directly from the
monotonicity of the logarithmic negativity under non-selective
trace-preserving completely positive maps. We wish to find a
PPT-map $\Psi$ that maps the maximally entangled state $\Phi(K_n)$
of $K_n$ dimensions to the target state $\rho^{\otimes n}$ for any
value of $n$, i.e., $\Psi(\Phi(K_n))=\rho^{\otimes n}$ for all
$n$. Then we have, for any $n$,
\begin{eqnarray*}
    \log_2 {\rm tr} |(\rho^{\Gamma})^{\otimes n}| &=& \log_2 {\rm tr}
    |(\Psi(\Phi(K_n)))^{\Gamma}|\\
    &\le& \log_2 {\rm tr} |\Phi(K_n)^{\Gamma}|= \log_2 K_n
\end{eqnarray*}
so that
\begin{displaymath}
    \log_2 {\rm tr} |\rho^{\Gamma}| \le \lim_{n\rightarrow\infty} \frac{1}{n}\log_2 K_n =
    E_{ppt}(\rho).
\end{displaymath}

Now we proceed to prove the upper bound on the entanglement cost.
The linear map $\Psi$ realising the transformation
$\Phi(K_n)\mapsto\rho^{\otimes n}$ must be completely positive and
trace preserving (CPTP), and PPT (which means that
$\Gamma\circ\Psi\circ\Gamma$ is positive as well). By proposing a
map that satisfies these criteria we directly find an upper bound
to the PPT entanglement cost. Consider thereto maps of the form
\beas \Psi(A) &=& aF+bG, \\ a &=& \trace(A\Phi(K_n)), \\ b &=&
\trace(A(\identity-\Phi(K_n))),
\\ F &=& \Psi(\Phi(K_n)), \\
G &=& \frac{\Psi(\identity-\Phi(K_n))}{K_n^2-1}. \eeas Note that
$a+b=1$ and $a,b\ge 0$. The requirements on $\Psi$ are that it
must be CPTP and PPT and must convert $\Phi(K_n)$ into the state
$\rho^{\otimes n}$. Thus $F=\rho^{\otimes n}$ and $G$ must be a
state. From the PPTness requirement,
$\Gamma\circ\Psi\circ\Gamma\ge 0$, it follows that $$ \forall A\ge
0: \trace(A\Phi(K_n)^\Gamma)F^\Gamma +
\trace(A(\identity-\Phi(K_n)^\Gamma))G^\Gamma \ge 0, $$ where have
made use of the self-duality of the partial transpose,
$\trace(X^\Gamma Y)=\trace(XY^\Gamma)$. Expressing the partial
transpose of $\Phi(K_n)$ in terms of the projectors on the
symmetric and antisymmetric subspaces ${\cal S}$ and ${\cal A}$,
respectively, \beas \Phi(K_n)^\Gamma &=& ({\cal S}-{\cal A})/K_n,
\\ \identity-\Phi(K_n)^\Gamma &=& (1+1/K_n){\cal A} +
(1-1/K_n){\cal S}, \eeas the PPTness condition becomes \beas
\forall A\ge 0: &&\trace(A{\cal A})(-F^\Gamma + (K_n+1)G^\Gamma) +
\\ &&\trace(A{\cal S})(F^\Gamma+(K_n-1)G^\Gamma) \ge 0. \eeas
Since ${\cal A}$ and ${\cal S}$ are mutually orthogonal projectors
and sum to the identity, this condition simplifies to the operator
inequality $$ -(K_n-1)G^\Gamma \le F^\Gamma \le (K_n+1)G^\Gamma.
$$ As a direct consequence, it follows that $G$ must be a PPT
state, $G^\Gamma\ge 0$, which was of course to be expected.

For PPT states, the PPT entanglement cost is obviously zero, so
that the optimal $K_n=1$. Therefore, we restrict ourselves in the
following to states $F=\rho^{\otimes n}$ that are not PPT; hence
$E_{ppt}>0$ and $K_n>1$. Obviously, $(1/n)\log K_n$ is a
non-increasing function of $n$, tending to $E_{ppt}>0$ in the
limit. Hence, for every $n$, $K_n\ge \exp(n E_{ppt})>1$. This
implies that, for every non-PPT state $\rho$, there is a number
$N$ such that $\forall n>N: K_n>>1$. For sufficiently large $n$,
therefore, the PPTness condition on the map $\Psi$ can be
approximated to arbitrary precision by the condition $-K_n
G^\Gamma \le F^\Gamma \le K_n G^\Gamma$.

We now propose to use the following state $G$, which incorporates
a correction term to ensure positivity of $G$:
\begin{eqnarray*}
G &=& \frac{(|\rho^\Gamma|^\Gamma + \alpha \identity)^{\otimes
n}}{Z^n}, \\ \alpha &=&
\max(0,-\lambda_{\text{min}}(|\rho^\Gamma|^\Gamma)),
 \\ Z &=& {\rm
tr}(\rho^\Gamma)+\alpha\dim(\rho).
\end{eqnarray*}
It is now easily seen that the PPTness condition for the map
$\Psi$ will be satisfied for the choice $K_n=Z^n$ (if $\alpha$ is
larger than zero, a somewhat smaller value of $K$ is possible but
we will not consider this possibility). Hence, we get the upper
bound for the PPT entanglement cost: $$ E_{ppt}(\rho) \le \log_2
Z(\rho). $$ \QED

In general, the lower and the upper bound in the theorem will not
coincide unless the bi-negativity is positive, i.e.\ if
$|\rho^{\Gamma}|^{\Gamma}\ge 0$. However, the vast majority of
quantum states have this property, as numerical investigations
indicate. Important examples for which the bi-negativity is
positive include the set of Werner states in $d\times
d$-dimensional systems, and all Gaussian bi-partite states in
infinite-dimensional systems with canonical degrees of freedom.
This will be proven in the subsequent two Lemmas.

\smallskip
\noindent
{\bf Lemma 1:} {\it Let $\rho$ be a Gaussian state defined on a
bi-partite system with a finite number of canonical degrees of freedom.
Then the bi-negativity satisfies $|\rho^\Gamma |^\Gamma\geq 0.$}

{\it Proof:} Let $\Gamma$ be the covariance matrix of $\rho$
\cite{Notation} and $P:=\text{diag}(1,...,1,1,-1,...,1,-1)$ be the
matrix corresponding to mirror reflection in one part of the
bi-partite system, i.e., partial transposition on the level of
states \cite{Notation}. Then, the normal mode decomposition
\cite{Vidal W02} (the Williamson normal form) of the covariance
matrix of $\rho^\Gamma$ can be written as $$
    S P \Gamma P S^T=:\text{diag}(x_1,x_1,...,x_{n}, x_{n}) ,
$$ with $x_i\geq 0$ for all $i=1,...,n$, where $S\in
Sp(2n,{\bf{R}})$ is an appropriate symplectic matrix. Therefore,
the problem of taking the absolute value has been reduced to an
effective single-mode problem. Going to the Fock state basis it is
then straightforward to see that the covariance matrix of
$|\rho^\Gamma|/ \|\rho^\Gamma \|_1$ is given by $$
    S^{-1}\left(S P \Gamma P S^T + p\right) (S^T)^{-1},
$$ where $p:=\text{diag}(p_1,p_1,...,p_n,p_n)$ is a positive
diagonal matrix with entries $$
    p_i= \left\{
    \begin{array}{ll}
    0, & \text{if $x_i\geq 1$,}\\
    1/x_i -x_i,  & \text{if $x_i<1$.}
    \end{array}
    \right.
$$ The state $\rho$ has a positive bi-negativity, i.e.,
$\rho_{bi}:=(|\rho^\Gamma|/ \| \rho^\Gamma\|_1)^\Gamma\geq 0$, iff
the covariance matrix $\Gamma_{bi}$ associated with $\rho_{bi}$
satisfies the Heisenberg uncertainty principle
$\Gamma_{bi}+i\Sigma\ge 0$ where $\Sigma$ is the symplectic matrix
\cite{Notation}. Hence, $\rho_{bi}$ is positive iff $$
    P S^{-1}\left(S P \Gamma P S^T + p\right) (S^T)^{-1} P + i \Sigma
    \geq 0.
$$ But as for the covariance matrix $\Gamma$ of the original state
$\rho$ we have $\Gamma + i \Sigma\geq 0$, and because $P S^{-1} p
(S^T)^{-1} P\ge 0$ this is indeed the case.\QED

\smallskip
 \noindent {\bf
Lemma 2:} {\it For any Werner state $\rho$ in a $d\times
d$-dimensional system the bi-negativity satisfies
$|\rho^{\Gamma}|^{\Gamma} \ge 0$.} \cite{Audenaert DVW02}

{\it Proof:} Any Werner state for a bi-partite state of two
$d$-dimensional subsystems can be written as
\begin{eqnarray*}
\rho &=& \frac{p(\openone - F)}{d(d-1)} + \frac{(1-p)(\openone + F)}{d(d+1)}\\
&=& q\openone + r |\Phi(d)\rangle\langle\Phi(d)|^{\Gamma}
\end{eqnarray*}
with $$
    q=\frac{p}{d(d-1)}+\frac{1-p}{d(d+1)},\,\,
    r=\frac{1-p}{d+1}-\frac{p}{d-1},
$$ and $F$ being the flip operator. Then we find
\begin{eqnarray*}
\rho^\Gamma &=& q(\openone - |\Phi(d)\rangle\langle\Phi(d)|) + (q+r) |\Phi(d)\rangle\langle\Phi(d)|
\end{eqnarray*}
and
\begin{eqnarray*}
|\rho^\Gamma|^{\Gamma} &=& q(\openone - \frac{F}{d}) +
\frac{|q+r|}{d} F.
\end{eqnarray*}
The eigenvalues of $F$ are $\pm 1$ and therefore the eigenvalues
of $|\rho^\Gamma|^{\Gamma}$ are easily checked to be non-negative.\QED

As a consequence, for Werner states, Gaussian states and for any
other states for which $|\rho^\Gamma|^{\Gamma}\ge 0$, such as pure
states \cite{Audenaert DVW02}, we have proven that the
entanglement cost for the exact preparation of the quantum state
$\rho$ using PPT-operations is given by the logarithmic
negativity.
This provides the, previously unknown,
operational interpretation of the logarithmic negativity for these
states. Note that the cost $E_{ppt}$ may generally coincide with
the logarithmic negativity even for states whose binegativity is
negative, but we were unable to prove or disprove this
possibility. Furthermore, note also the surprising fact that the
PPT-cost for exact preparation is a concave function on Werner
states (see also Fig 1). This implies, rather counter-intuitively,
that mixing, i.e.\ the loss of information, may increase the
PPT-cost for exact preparation.
We proceed by using the Theorem together with Lemma 2 to provide a
result on the PPT-entanglement cost for the anti-symmetric Werner
state.

\begin{figure}[hbt]
\begin{center}
    \epsfxsize=8.5cm
       \epsfbox{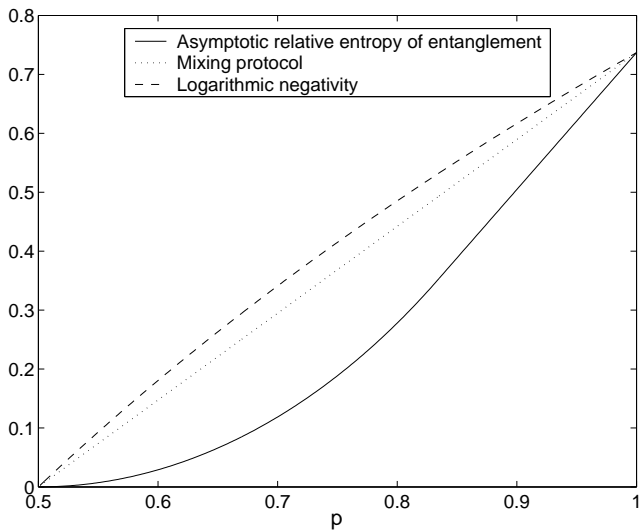}
\end{center}
\vspace*{-0.5cm} \caption{ Various entanglement costs plotted for
Werner states of the form $\rho=p\sigma_a+(1-p)\sigma_s$ for
qutrits. The dashed line represents the PPT-cost for the exact
preparation of $\rho$ which is a concave function. The dotted line
is the mixing protocol (One can create asymptotically the Werner
state $\rho=(2p-1)\sigma_a+(1-p)(\sigma_a+\sigma_s)$ by creating
the state $\sigma_a$ with probability $2p-1$ and the state
$\sigma_a+\sigma_s$ with probability $(1-p)/2$ at a cost given by
the straight line), and represents the best known PPT-cost for
approximate but asymptotically exact preparation of $\rho$. The
solid line is the asymptotic relative entropy of entanglement for
PPT-operations.}
\end{figure}

\smallskip
\noindent
{\bf Lemma 3:} {\it The PPT-entanglement cost $C_{ppt}$ for the
anti-symmetric Werner state $\rho=\sigma_a$ is given by
$LN(\rho)$ and coincides with its PPT-distillable entanglement
$D_{ppt}(\rho)$.}

{\it Proof:} From Lemma $2$ we know that the bi-negativity of
$\sigma_{a}$ is positive. As a consequence from the Theorem we
conclude that
$E_{ppt}(\rho)=LN(\rho)$. This provides an upper bound on the
entanglement cost for asymptotically exact preparation of the
states, i.e., $LN(\rho)=E_{ppt}(\rho)\ge C_{ppt}(\rho)$. On the
other hand a lower bound is given by the PPT-distillable
entanglement of $\sigma_a$, which has been computed in \cite{Rains
00} and which equals the logarithmic negativity as well. Therefore
we have $LN(\rho)=E_{ppt}(\rho)\ge C_{ppt}(\rho)\ge
D_{ppt}(\rho)=LN(\rho)$ and all the quantities coincide. \QED

This Lemma is remarkable, as it shows that asymptotic entanglement
transformations can be reversible even for truly mixed states, as
long as one considers the class of PPT-operations. The result of
Lemma 3 may still be a coincidence as it refers to an extreme
point of a set of states (here the set of $U\otimes U$ symmetric
states, i.e., the Werner states) but further evidence from
numerical studies suggest that PPT-entanglement cost and
PPT-distillable entanglement converge towards each other on Werner
states. This collection of evidence makes it plausible to ask the
question as to whether the entanglement cost under PPT-operations
coincides with the PPT-entanglement of distillation or, in other
words, whether asymptotic entanglement transformations are reversible
under PPT-operations. If the answer to this question would be
affirmative this would simplify the theory of quantum entanglement
considerably. This would furthermore indicate that the theory
of mixed state entanglement takes its most elegant form in the
framework of PPT-operations. This and the other results in this
work reveal that PPT-operations are a most useful concept for the
study of quantum entanglement, meriting further investigations into
their properties.

This work was supported by EPSRC, the European Union EQUIP
project, the ESF Programme on 'Quantum Information Theory and
Quantum Computing' and by a Feodor-Lynen grant of the
Alexander-von-Humboldt Foundation. We thank P.\ Horodecki, S.\
Virmani, K.G.H.\ Vollbrecht and R.F.\ Werner for discussions.


\begin{thebibliography}{99}
\bibitem{Notation}
The entries of the covariance matrix $\Gamma$ are given by
$\Gamma_{jk}= 2\text{tr}[\rho( O_j - \langle O_j\rangle_\rho) ( O_k - \langle O_k\rangle\rho) ]
- i \Sigma_{jk}$, for $j,k=1,...,2n$,
where $(O_1,...,O_{2n})=(X_1,P_1,...,X_n,P_n)$
is the vector collecting the canonical coordinates of this system
with $n$ degrees of freedom, and $\Sigma$ is the symplectic matrix
incorporating the canonical commutation relations.
We follow the conventions in Refs.\ \cite{Gauss}.
\end{thebibliography}
\end{document}